# Physics-based material parameters extraction from perovskite experiments via Bayesian optimization

Hualin Zhan[1*], Viqar Ahmad[1], Azul Mayon[1], Grace Tabi[1], Anh Dinh Bui[1], Zhuofeng Li[1], Daniel Walter[1], Hieu Nguyen[1], Klaus Weber[1], Thomas White[1*], and Kylie Catchpole[1*]

[1]*School of Engineering, Australian National University, ACT 2601, Australia*

Corresponding Authors
*Hualin Zhan: hualin.zhan@anu.edu.au
*Thomas White: tpwhitehome@gmail.com
*Kylie Catchpole: kylie.catchpole@anu.edu.au

**Abstract**

**The ability to extract material parameters of perovskite from quantitative experimental analysis is essential for rational design of photovoltaic and optoelectronic applications. However, the difficulty of this analysis increases significantly with the complexity of the theoretical model and the number of material parameters for perovskite. Here we use Bayesian optimization to develop an analysis platform that can extract up to 8 fundamental material parameters of an organometallic perovskite semiconductor from a transient photoluminescence experiment, based on a complex full physics model that includes drift-diffusion of carriers and dynamic defect occupation. An example study of thermal degradation reveals that the carrier mobility and trap-assisted recombination coefficient are reduced noticeably, while the defect energy level remains nearly unchanged. The reduced carrier mobility can dominate the overall effect on thermal degradation of perovskite solar cells by reducing the fill factor, despite the opposite effect of the reduced trap-assisted recombination coefficient on increasing the fill factor. In future, this platform can be conveniently applied to other experiments or to combinations of experiments, accelerating materials discovery and optimization of semiconductor materials for photovoltaics and other applications.**

**Introduction**

Organometallic perovskites are a promising class of semiconductor that have led to excellent efficiency in solar cell and lighting research. Despite the rapid progress, perovskites exhibit low stability and large variations in their performance among different labs, presenting challenges in commercialisation. Our ability to accurately measure the fundamental material properties of perovskites such as electron/hole mobility, dopant concentration, defect energy level, etc. is the key to overcoming these challenges. However, measured values of, for example, carrier mobility, can vary across three orders of magnitude for a common perovskite of methylammonium lead iodide[1-3], indicating that accurate and low-cost measurements of fundamental properties of perovskites are difficult and not widely available. Such measurements involve extracting material properties using analysis methods that should ideally be accurate, physically interpretable, rapid, low-cost, and adaptable to a range of experiments. However, conventional accurate and rapid experimental analysis methods developed for silicon or III-V semiconductors, e.g., extraction of dopant concentration and carrier mobility from low-cost measurement of resistivity, typically rely on simplified empirical models that are derived under specific conditions, which fail to describe perovskites[4, 5].

In perovskite research, various techniques including space-charge-limited current, drive-level capacitance profiling, and microwave and photoluminescence (PL) experiments are used to extract the carrier mobility, defect energy level and recombination coefficients, respectively[1, 2, 6, 7]. The often simplified, conventional models used to interpret these measurements have raised questions about the accuracy of inferred parameter values[8-10], suggesting more detailed models such as the drift-diffusion model of carriers should be used due to the presence of electric fields and concentration gradients[8]. However, accurate analysis/parameter extraction based on the drift-diffusion model has been a significant challenge due to the difficulty in fitting the solution of a large set of partial differential equations to experiments. If we include additional physics, e.g., the dynamic model of defect occupancy that may be required for shallow defects[11], the number of equations increases further. To date, experimental analysis based on fundamental physics models has been qualitative or semi-quantitative at best, i.e., identifying the range of values for the parameters of interest[12, 13]. This challenge significantly hinders the rational development of novel materials.

Here, we focus on a widely used PL-based technique – Time-resolved photoluminescence (TRPL) – because, in theory, it can quantify the fundamental properties that determine the optoelectronic quality of semiconductors including carrier mobility, defect energy level, etc. In practice, however, this quantification is difficult for perovskites due to the complication of interpretation using different models[14-16]. TRPL studies the time-dependent recombination of photo-generated carriers via various processes. Conventional analysis of TRPL based on straightforward exponential fitting to the experimental data gives a composite parameter called carrier lifetime, which describes the time before most of the photo-generated carriers recombine[16, 17]. More sophisticated analyses of TRPL have employed generic kinetic models that describe the temporal evolution of carriers by considering their photo-excited generation and parameterized recombination processes, from which composite parameters such as rate constants can be extracted[18, 19]. However, neither of these approaches quantitatively extracts the fundamental material properties, such as the defect energy level and the individual electron/hole mobility, which are needed for input into semiconductor device models[13, 15].

Machine learning (ML) has emerged recently as a viable technique to assist semiconductor characterization, in which material properties are generally extracted from new data-based models that are trained by a massive theoretical/physical dataset[20-24]. For TRPL characterization, researchers have built a massive database for ML and can extract at least 4 material parameters from TRPL measurements on a single perovskite material[22], which includes 2 fundamental parameters, i.e., radiative recombination coefficient and equilibrium hole concentration, and 2 composite parameters, i.e., ambipolar carrier mobility (a function of individual electron/hole mobility) and non-radiative carrier lifetime. This approach heavily relies on the dataset production, which requires high-performance computation and re-production if researchers change theoretical models, materials structures, or experiments, leading to difficulty for use in everyday research.

Here, without the need to produce a massive theoretical database, we use ML, and specifically Bayesian optimization (BO), to develop a material analysis platform capable of extracting up to 8 fundamental material parameters, such as carrier mobility, defect energy level, etc., from TRPL experiments on a single perovskite material. This platform adopts a fundamental full-physics model that includes the drift-diffusion of carriers, radiative recombination, Auger recombination, and the dynamic occupation of defects. We first demonstrate how ML should be combined with our understanding of the physics to design a new cost function for efficient parameter extraction, rather than being used in isolation. Then we discuss strategies for reliable ML-assisted extraction of multiple parameters by further exploiting our physical understanding. Lastly, we demonstrate a case study of using this platform to reveal the change in perovskite properties during thermal degradation, a topic of great interest to perovskite development. Beyond TRPL, the database-independence allows this platform to be conveniently applied to other experiments or to combinations of experiments.

**Material parameters analysis by Bayesian optimization**

Material parameters are generally used in theoretical simulation to predict the experimental behaviour of a material or device. If the prediction is quantitatively correct, the parameters used are considered adequate to describe the materials. As such, material parameters can be extracted from experiments via a trial-and-error search process, in which different values of the material parameters are tested to achieve a quantitative comparison/best fit between the experiment and simulation. A simple grid search, i.e., testing all possible combinations of parameter values, may be possible for two parameters[25], but becomes unmanageable when the number of parameters increases and the potential parametric values span several orders of magnitude, e.g., a grid search of 5 parameters, each spanning 5 orders of magnitude, would require $(5×4)^5$=3.2 million simulation runs to sample 4 values in each order of magnitude. A TRPL simulation in this work takes around 17 seconds on a laptop computer (Apple M1), leading to an unfeasible total simulation time of 15 thousand hours for the grid research.

Here, we use BO to accelerate the search process, so that only the parametric values with high probabilities of fitting the experiment are tested. BO guides the parameter-search iteratively: In each iteration, BO finds (most probably) the "best" parameters for simulation given the information it has to date, and the simulation then returns its experiment-fitting quality to BO for the next iteration (Fig. 1a). In this way, BO is iteratively updated/trained for accelerated search using a comparison between the experiment and the previously simulated data (blue box). Our BO-guided analysis platform takes the experimental data as input and outputs all tested parameters including those that best fit the experiment. A BO search in this work commonly takes 600 iterations, i.e., around 2 hours and 50 minutes in the same Apple M1 laptop. The run time of BO is similar to that of two other optimization methods – a gradient-based method and simulated annealing – discussed later and is far less than another commonly used method called genetic algorithm, which generally takes about hundreds of thousands of iterations[26]. This makes BO a suitable choice for this work.

This platform has two immediate advantages: (a) high flexibility – the separate roles of BO in parameter-search guidance, and simulation in physics-based prediction, enable the freedom to use any theory of choice and hence any experiment of interest; and (b) no need for a pre-produced massive dataset. The benefits are twofold. Without the need to re-create a dataset, researchers can conveniently modify the studied material structure and the physics, e.g., including additional material layers and/or introducing ion transport theory. Furthermore, the parameter search capability of pre-produced datasets is dependent on the resolution in the parameter space, i.e., the number of parametric values tested in each order of magnitude. In contrast, the dataset in this platform is produced during, rather than before, parameter search, and thus has much higher resolution around the most probable parametric values than elsewhere, which increases the search efficiency and decreases the computational demand. These benefits enable straightforward implementation of this platform on a desktop computer for customized materials, characterization, and physical models.

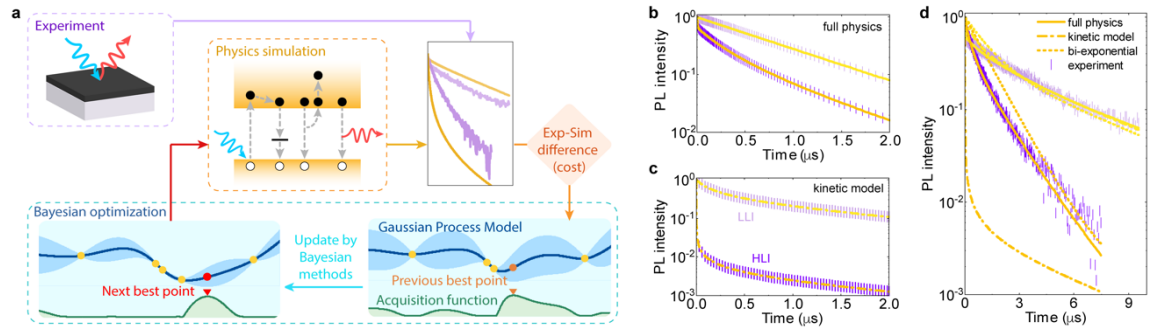

**Fig. 1 | Material parameters analysis platform and preliminary validation. a,** Flow chart of the platform. Bayesian optimization (cyan box) iteratively provides the best estimation of parameters for the physics simulation (yellow box) based on the comparison between the simulation and experiments (purple) box from the last iteration. Inside the cyan panel: the Gaussian Process Model (blue sausage-shaped curve) plots the possible cost vs. parameter value; the acquisition function (green curve) estimates the likelihood of the best-fitting parameter value. **b,c,** The platform can successfully find the best the fit (yellow) to a previously numerically simulated target TRPL (purple) for different choices of physical model, i.e., full-physics model (**b**) and the generic kinetic model (**c**). The upper and lower curves in each case are for low-level injection (LLI) and high-level injection (HLI) conditions of TRPL. **d,** Preliminary experiment analyses by the platform using various models, showing that the full physics model is required to fit the experiment. In each analysis, all parameters are the same for LLI and HLI.

We first demonstrate, in preliminary cases, the validity and flexibility of this platform and the necessity of the full-physics model for TRPL analysis of perovskite. We use the full-physics model [Eq. (1)-(12)] and the generic kinetic model [Eq. (13)] to compute target TRPL curves for analysis (purple symbols in Fig. 1b and c). The platform finds the best fits (yellow lines) and the parameters using the corresponding model of the targets, confirming the validity of the platform and the flexibility to apply different theoretical models. However, when experiments are the target (measured on a perovskite/glass structure), this platform only finds the best fits using the full-physics model (Fig. 1d), confirming the necessity of this model for TRPL analysis.

**Straightforward extraction of selected parameters**

In this platform (Fig. 1a), the difference between the simulation and the experiments is quantified by a cost function (orange diamond) in every iteration and therefore, the technical objective is to minimize the cost. Rather than testing all parametric values in simulation and looking for the minimum, BO constructs mathematical relations between the cost and the parameters using the Gaussian Process Model, which assumes that all possible relations follow a Gaussian probability distribution[27]. The distribution is updated after every simulation by the calculated cost based on the Bayesian method, which is then translated to a function called the acquisition function to predict the next parametric value (red symbols) with the highest probability to best fit. Thus, the cost function plays a vital role in predicting the next best point.

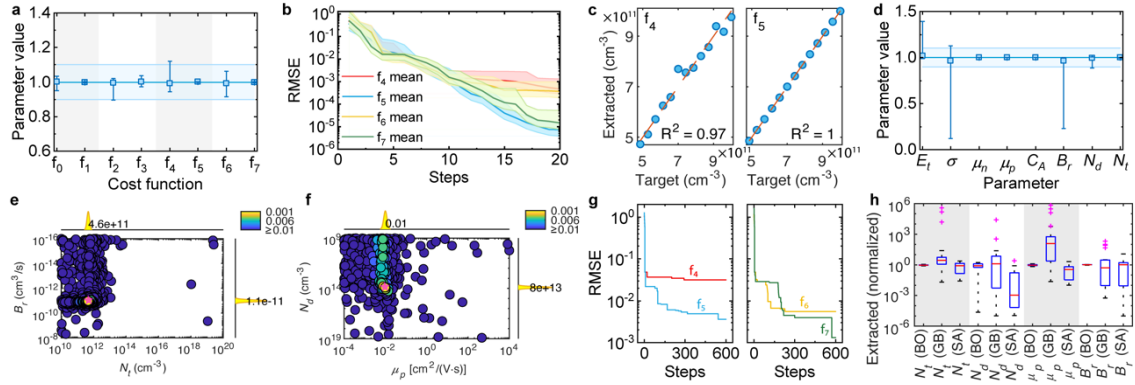

**Fig. 2 | Fitting performance and straightforward implementation of BO for parameter extractions. a,** Single-parameter extraction of $N_t$ by different cost functions. **b,** Convergence of the extraction of $N_t$ by different cost functions. **c,** Cross-validation plots for the extraction of $N_t$ by $f_4$ and $f_5$. **d,** Single-parameter extraction of various parameters by $f_7$. In (**a**) and (**d**), all extracted parameters are normalized by their target values. The error bars are the maximum and minimum extracted parameters from 22 repeated fittings. The blue squares are the mean extracted parameter. **e,f,** Four-parameter extraction of $B_r$, $N_t$, $N_d$, $\mu_p$. The circles represent all tested values in the 4D parametric space projected onto two 2D planes of ($B_r$, $N_t$) and ($N_d$, $\mu_p$). The colour scale represents the cost, where the yellow points indicate the lowest cost and hence give the extracted parameter. The target values are indicated by the magenta star: $B_r=10^{-11}$ cm$^3$/s, $N_t=5\times10^{11}$ cm$^{-3}$, $N_d=10^{14}$ cm$^{-3}$, $\mu_p=0.01$ cm$^2$/(V·s). **g,** Convergence of the four-parameter extraction of $B_r$, $N_t$, $N_d$, $\mu_p$ by different cost functions. In (**b**) and (**g**), all differences between the target and the simulation are translated to RMSE using $f_6$ so that they can be compared on the same basis. **h,** Four-parameter extraction by the methods of BO, a gradient-based method (GB), and simulated annealing (SA). $f_5$ is used for all methods. Each method is repeated 22 times. The red lines are the median, the blue boxes indicate the 25$^{th}$ and 75$^{th}$ percentile, the black dashed lines are the non-outlier minimum and maximum, and the magenta pluses are the outliers. All parameter extractions are based on fitting to TRPL under both LLI and HLI conditions.

We design eight cost functions [Eq. (14)-(21)] and categorize them into two types – the RMSE (root-mean-square error) type ($f_0, f_2, f_4, f_6$) and their logarithms ($f_1, f_3, f_5, f_7$), e.g., $f_6 = \sqrt{\sum_i (h_i - y_i)^2}$ and $f_7 = \log\left[\sqrt{\sum_i (h_i - y_i)^2}\right]$, where $h_i$ and $y_i$ are the simulated and experimental/target TRPL at the time point $i$. Fig. 2a compares the extraction of defect density ($N_t$) from a space between $10^{10}$ and $10^{20}$ cm$^{-3}$ (denoted as [$10^{10}$, $10^{20}$] cm$^{-3}$) within 20 steps/tests using 8 cost functions. Here, the values of all other parameters are known, i.e., parameter-search is performed in one-dimensional (1D) parametric space. Although $N_t$ is correctly extracted using all cost functions, the logarithm-type cost functions finding the target more accurately than their RMSE-type counterparts, e.g., $f_5$ vs. $f_4$, as demonstrated by the smaller error bar (Fig. 2a), the lower cost (Fig. 2b), and the better $R^2$ value (Fig. 2c).

Further test of 1D parameter-search successfully extracts seven other parameters (Fig. 2d): relative defect energy level ($E_t$) from [0, 0.5], capture cross-section for trap-assisted recombination ($\sigma$) from [$10^{-20}$, $10^{-12}$] cm$^2$, electron and hole mobility ($\mu_n$ and $\mu_p$) from [$10^{-4}$, $10^4$] cm$^2$/(V·s), Auger recombination coefficient ($C_A$) from [$10^{-32}$, $10^{-27}$] cm$^6$/s, radiative recombination coefficient ($B_r$) from [$10^{-16}$, $10^{-8}$] cm$^3$/s, and dopant density ($N_d$) from [$10^9$, $10^{19}$] cm$^{-3}$. Here $E_t$ represents the distance between the actual defect energy level and the conduction or valance band edge (whichever is smaller) divided by the band gap. This indicates that each of these parameters has a unique solution in TRPL fitting in this study, i.e., fundamentally extractable when one parameter is unknown.

We then increase the number of the unknown parameters and test a 4D parameter-search where $B_r$, $N_t$, $N_d$, and $\mu_p$ are unknown (Fig. 2e, f). Again, the extracted parameters are close to their targets. The logarithm-type cost functions converge faster and extract better parametric values than their RMSE-type counterparts (Fig. 2g). We compare the performance of BO in 4D parameter-search with two other optimization methods – a gradient-based method and a Monte-Carlo method called simulated annealing. We repeat 22 times for each method and find that BO outperforms two other methods: all four parameters extracted by BO have median values close to the target and the smallest deviation among all methods (Fig. 2h). We note that, even with BO, the types of parameters that can be extracted in 4D parameter-search is limited to a few selected combinations. Changing a parameter in these combinations or including additional parameters can lead to a large uncertainty in the extracted parameters when directly using BO for multi-parameter search (Supplementary Fig. S1, Fig. S2).

## Multi-parameter extraction via a physics-based strategy

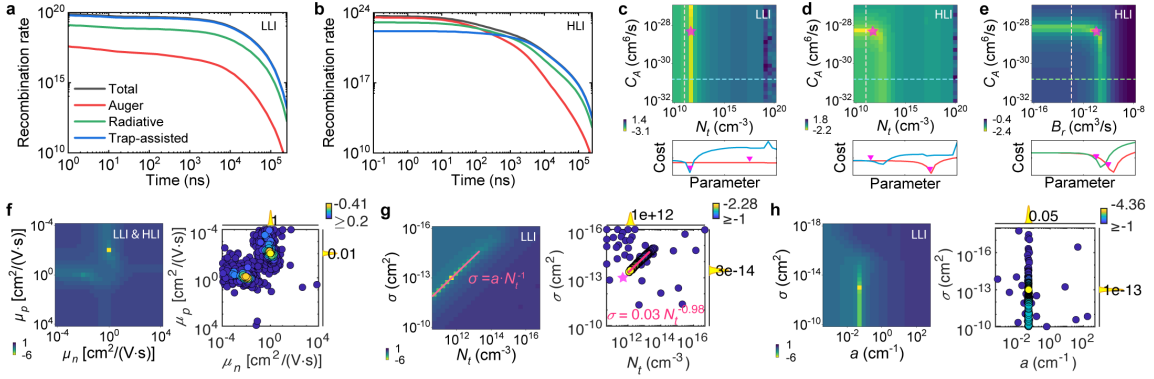

**Fig. 3 | Translation of the underlying physics to a data-based perspective using simulated results. a,b,** Total electron recombination rate contributed by different recombination mechanisms under LLI (**a**) and HLI (**b**). **c-e,** 2D grid searches for $N_t$ & $C_A$ (**c,d**) and $B_r$ & $C_A$ (**e**), where all the rest parameters are fixed. Bottom panels: cost-value profile along the dashed lines where the x-axis is normalized. The magenta stars indicate the target values, and the triangles are their projection on each axis. **f-h,** 2D grid (left panels) and BO (right panels) searches for $\mu_n$ & $\mu_p$ (**f**), $N_t$ & $\sigma$ (**g**), and $a$ & $\sigma$ (**h**). While grid search tests all parameter values, BO focus on testing the parameter values that are likely to give the best fit. The pink line in the left panel of (**g**) shows the correlation between $N_t$ and $\sigma$, where the correct value of $a$ is 0.05. The red line in the right panel of (**g**) is a fit to the direct search results of $N_t$ and $\sigma$ by BO, where $a$=0.03. Including the relation $\sigma = a/N_t$ helps BO to find the correct values (**h**).

Here, we demonstrate that, when combined with our understanding of the mechanisms behind TRPL measurement, we can devise a physics-based fitting strategy for use in the ML workflow, so that we can extract eight unknown parameters. We note that the trap-assisted recombination mechanism dominates TRPL under the condition of low-level injection (LLI), whereas Auger recombination dominates under high-level injection (HLI) at the beginning (Fig. 3a, b).

These different recombination mechanisms can be translated into a data-based perspective to provide an indication of the intrinsic limitations and opportunities for fitting. Fig. 4c-e show 2D grid searches for the parameters that influence recombination. The yellow points in Fig. 3c form a line parallel to the y-axis, suggesting the extraction of $N_t$ is independent of $C_A$, i.e., $N_t$ can be extracted for any value of $C_A$. Indeed, while the minimum of the cost profile of $N_t$ along the cyan dashed line, i.e., the blue curve in the bottom panel, coincides with its target value (purple triangle), the cost profile of $C_A$ remains nearly constant, indicating that $N_t$ is easier to extract than $C_A$ under LLI. In contrast, $C_A$ is easier to extract than $N_t$ under HLI (Fig. 3d). Here, the extraction of $N_t$ (or $C_A$) is nearly independent of $C_A$ (or small $N_t$) under LLI (or HLI).

This observation suggests two methods for fitting. (a) Straightforward fitting to both LLI and HLI conditions as used in Fig. 2. An additional example is the inclusion of the LLI condition in the extraction of $\mu_n$ & $\mu_p$, which successfully identifies the individual value of $\mu_n$ & $\mu_p$ (Fig. 3f) rather than the ambipolar mobility – an observation that is consistent with electrical experiments[28]. Note that the ambipolar mobility could be extracted when considering HLI-only experiment and using $f_6$ (Supplementary Fig. S3 and ref[22]). However, as discussed earlier, fitting to both LLI and HLI becomes difficult when the number of parameters increases. (b) Separating the extraction of the parameters that are independent of each other. In addition to above example of separately extracting $N_t$ & $C_A$, $B_r$ and $C_A$ independently dominate the radiative and Auger recombination and can be estimated separately by artificially lowering $C_A$ and $B_r$ to reduce the effect of Auger and radiative recombination, respectively. Note that separate fitting estimates the range of the studied parameters, rather than giving their exact values (e.g., minima in the green and red curves vs the purple triangles in Fig. 3e). The estimated values can be used to define a smaller parametric range/space for more effective multi-parameter fitting. Although separate fitting might sound unnecessary when extracting only a few parameters, it significantly improves the extraction of multiple parameters because the ability to extract one parameter can be affected by its interaction/correlation with other parameters (Supplementary Note).

For the parameters that are correlated with each other and cannot be separated directly by different recombination mechanisms, we can re-construct the parameters and reduce the correlation. Here, we investigate an example, $N_t$ & $\sigma$, which are correlated in trap-assisted recombination via $\sigma = a/N_t$ [Eq. (14)-(16)]. The trap-assisted recombination rate is proportional to $a$, which becomes a constant if the second terms in Eq. (14) and (15) are negligible. This condition is approximately valid when $\sigma$ is large and hence, the trap-assisted recombination is predominantly affected by $a$ for any choice of $N_t$ & $\sigma$, presenting challenges in extracting $N_t$ and $\sigma$ (Fig. 3g). As a result, extracting $a$ becomes easier (Fig. 3h). The extraction of $\sigma$ is still possible under the LLI condition, although the uncertainty could be large because the effect of $\sigma$ on trap-assisted recombination is small [Eq. (14)-(16)]. The embedded physics-based relation $\sigma = a/N_t$ accelerates BO in finding the correct parameter values by skipping the construction of data-based relation between $N_t$ & $\sigma$.

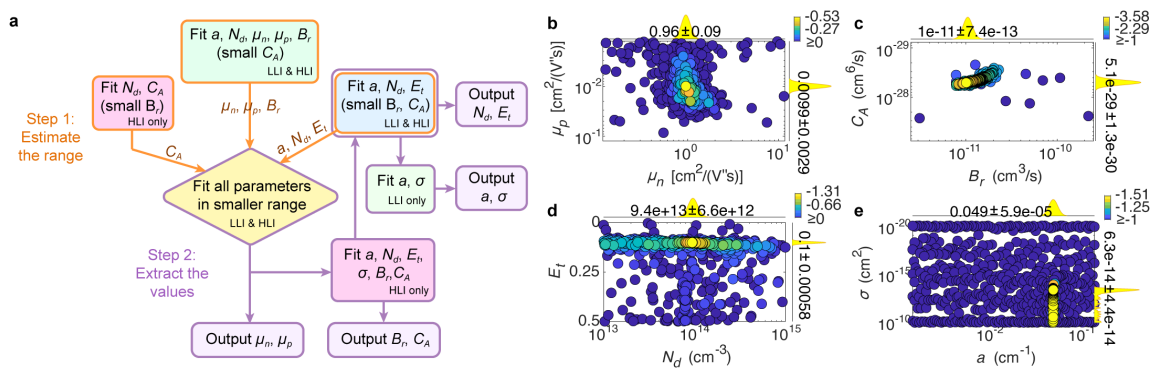

**Fig. 4 | Multi-parameter extraction. a**, The extraction strategy, which utilises separate fitting. The orange and purple arrows indicate the flow of step 1 and 2, respectively. **b-e**, 8D material parameter extraction from previously simulated TRPL based on a perovskite model with the bandgap of 1.68 eV. The target values are $\mu_p$=0.01 cm$^2$/(V·s), $\mu_n$=1 cm$^2$/(V·s), $C_A$=5×10$^{-29}$ cm$^6$/s, $B_r$=10$^{-11}$ cm$^3$/s, $N_d$=10$^{14}$ cm$^{-3}$, $E_t$ =0.1(×1.68 eV)=0.168 eV, $a$=0.05 cm$^{-1}$, $\sigma$=1×10$^{-13}$ cm$^2$.

Separate fitting allows us to develop a fitting strategy for multi-parameter extraction which involves two general steps (Fig. 4a): (1) reduce/estimate the parametric range and (2) improve the fitting. In step 1, we use HLI-only fitting to estimate $C_A$. Simultaneous fitting to HLI and LLI is used for estimating $B_r$, $\mu_n$, $\mu_p$, $N_t$, and $E_t$. The conditions in the parentheses are used in three separate fittings in step 1 only to improve the estimations. We set $E_t$=0.5 and $\sigma$=10$^{-11}$ cm$^{-2}$ when not specified because these choices marginally affect the fitting at this stage [Eq. (14)-(16) and Fig. 3h]. In step 2, we first extract $\mu_n$ and $\mu_p$. Note that a local optimum (a set of parameter values that is not the target but produces a similar TRPL) of $\mu_n$ & $\mu_p$ is observed symmetric to the target (yellow and orange points in Fig. 3f, see Supplementary Fig. S3 for more). Step 1 could output this local optimum during estimation. In the following extraction of $C_A$, $B_r$, $N_d$, $E_t$, $N_t$ and $\sigma$, we use two sets of values of $\mu_n$ & $\mu_p$ that are symmetric to each other, e.g., ($\mu_n$=0.01, $\mu_p$=1) and ($\mu_n$=1, $\mu_p$=0.01), and compare the costs. If they are significantly different, we take the extracted values that give the smallest cost.

All these physics-based strategies, including separate fitting, reducing the parametric correlation, reducing the parametric range, and taking the symmetric values of estimated $\mu_n$ & $\mu_p$, can be summarized into an effort of assisting BO to effectively find the target (or global optimum) from a group of local optima. We observe that BO is effective in finding the global optimum under two conditions, which can be approached by these strategies: (a) the local optima form continuous lines parallel to the axes of the map; and (b) the global optimum does not appear as a singularity, i.e., a point where the cost value changes drastically in a very small region. With these strategies, our BO-enabled platform successfully extracts all 8 parameters in examples where the role of trap-assisted recombination varies (Fig. 4b-e, Table 1, and Supplementary Fig. S4-S6).

**Table 1 | Extracted material parameters of perovskite from a simulated TRPL result.**

|  | Validation 1 | | Validation 2 | |
| --- | --- | --- | --- | --- |
|  | Target | Extracted | Target | Extracted |
| Electron mobility [cm$^2$/(V·s)] | 1 | 0.96±0.09 | 1 | 1±0.0076 |
| Hole mobility [cm$^2$/(V·s)] | 0.01 | 0.0099±0.0029 | 1 | 1±0.0076 |
| Radiative recombination coefficient (cm$^3$/s) | 1×10$^{-11}$ | (1±0.07) ×10$^{-11}$ | 1×10$^{-11}$ | (1.06±0.001) ×10$^{-11}$ |
| Auger recombination coefficient (cm$^6$/s) | 5×10$^{-29}$ | (5.1±0.1) ×10$^{-29}$ | 6×10$^{-30}$ | (4.7±0.06) ×10$^{-30}$ |
| Dopant density(cm$^{-3}$) | 1×10$^{14}$ | (9.4±0.7) ×10$^{13}$ | 1×10$^{14}$ | (9.7±1.9) ×10$^{13}$ |
| Defect energy level (away from band edge) | 0.1 (×1.6186 eV) ≈ 0.16 eV | 0.1±0.0006 (≈ 0.16 eV) | 0.0625 (≈ 0.1 eV) | 0.0608±0.0003 (≈ 0.098 eV) |
| Capture cross-section: defect-assisted (cm$^2$) | 1×10$^{-13}$ | (6.3±4.4) ×10$^{-14}$ | 2×10$^{-17}$ | (1.8±0.06) ×10$^{-17}$ |
| $a$ (cm$^{-1}$) [or Trap density (cm$^{-3}$)] | 0.05 (or 5×10$^{11}$) | 0.049±6×10$^{-5}$ (or 7.8×10$^{11}$) | 0.02 (or 1×10$^{15}$) | 0.017±1×10$^{-4}$ (or 9.41×10$^{14}$) |

**Changes in perovskite material parameters during thermal degradation**

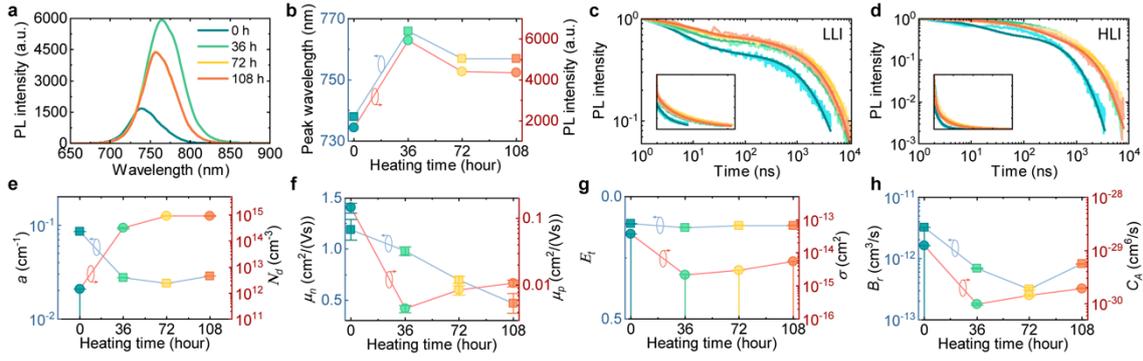

**Fig. 5 | Change of the material properties of perovskite during thermal degradation. a,** Steady-state PL measurement of perovskite films on glass substrate after heating in nitrogen at 85 ºC for 0h, 36h, 72h, 108h. **b,** Variation of the peak wavelength and PL intensity over the heating time. **c,d,** TRPL measurements under LLI (**c**) and HLI (**d**) conditions. **e-h,** 8D material parameter extraction from (**c**) and (**d**).

We utilise this platform to study how the material parameters change for perovskite under heating. Here, we fabricate a triple cation perovskite ($Cs_{0.21}FA_{0.74}MA_{0.05}PbCl_{0.11}Br_{0.43}I_{2.46}$) film on glass and place the sample in nitrogen at 85 ºC for different periods of time – 0h, 36h, 72h, and 108h. The hyperspectral steady-state PL measurement exhibits changes in both the peak wavelength and intensity after heating (Fig. 5a, b), suggesting a change in the material. Our step-by-step fitting leads to good agreements between theory and TRPL measurements under both LLI and HLI conditions for all heating times (Fig. 5c, d, see Supplementary Fig. S7, S8 for details). Different peak wavelength in Fig. 5a is converted to different bandgap used as a known parameter in the TRPL fitting. The parameters extracted from fittings are the same for LLI and HLI, but different for each heating time.

We confirm the material change by observing a change in $N_d$ and $B_r$, which are considered constant for unchanged materials. We find that heating in nitrogen reduces the trap-assisted recombination and the carrier mobility. Note that, due to the difficulty in distinguishing electron mobility from hole mobility in this case, the observed change in mobility could be symmetric (Supplementary Fig. S7, S8). Further experiments, such as electrical measurements which are mostly affected by the majority carrier, could be included in this platform in future to identify the type of carrier. The obtained $E_t$ indicates shallow defects in perovskite, which is consistent with past discoveries[29]. Further, we find heating in nitrogen has small effect on $E_t$. $C_A$ and $\sigma$ also decrease after heating, suggesting suppressed non-radiative recombination, but the associated uncertainty can be large. In future, we can utilise long-term TRPL measurements, such as the TRPL stitching technique[16], and/or test different models, such as considering different Auger recombination coefficients for electrons and holes, to reduce the uncertainty.

The extracted material parameters enable us to discuss the possible effect of heating on the performance of perovskite-based solar cell. We find that the fill factor in *J-V* measurements can be reduced by heating (Supplementary Fig. S9), which is a competing effect from mobility and trap-assisted recombination coefficients (*a*). In general, small recombination leads to high fill factor because of the small series resistance. While the reduced recombination coefficients directly decrease the recombination, which could induce high fill factor (dashed line in Fig. S9a), the reduced mobility allows for carriers to stay longer in the perovskite before extraction to the transport layer, which increases the chance for carrier recombination and hence substantially reduces the fill factor (dotted line in Fig. S9a). However, we must note that, in this discussion we neglect the contribution from mobile ions, omit interface recombination, exclude passivation layers, and have made assumptions on the properties of the transport layers (Supplementary Table 2), which all affect the performance of solar cells. In future, we can use our platform to analyse these properties for more comprehensive studies, which could require additional characterization techniques. Supplementary Fig. S10 shows an example that our platform can be used for different experiments, not only further validating the extracted parameter but also providing rich prospects for future studies.

**Conclusion**

In this work, we have developed an effective material parameters extraction platform by adapting a BO algorithm in theory-experiment fitting processes, in which new logarithm-type cost functions and physics-based fitting strategies are used. The advantages of this platform include (1) freedom in selecting/modifying any theory, material model, and/or experiment so that it is highly flexible, and (2) no need to create a database which allows for implementation on desktop computers. These advantages allow us to extract eight fundamental properties of perovskites from a full-physics model that incorporates the dynamics of carriers and defect occupation for TRPL analyses. In an example study of thermal effects on perovskite, we find that long-time heating in nitrogen can increase the dopant density and decrease the carrier mobility, while most of the other properties change negligibly. In future, this approach could be readily applied to other fields. Although a direction implementation of BO can extract a few selected parameters, we recommend to develop corresponding physics-based fitting strategies based on the domain knowledge of these fields to expand the capability of BO.

**Methods**

**Fabrication of the triple cation perovskite sample.** The perovskite precursor (2 mL, 1.2 M) is prepared by dissolving CsI (137.2 mg, Sigma), FAI (322 mg, GreatCell), MACl (8.1 mg, Dynamo), $PbCl_2$ (20 mg, Sigma), $PbBr_2$ (198.2 mg, Sigma), and $PbI_2$ (857.5 mg, TCI) in a 1.6 mL of N,N-dimethylformamide (DMF) and 0.4 mL of dimethyl sulfoxide (DMSO). Perovskite is deposited on the glass substrate by spinning the precursor solution at 3000 rpm for 12 s (acceleration rate 100 rpm/s). The sample is then transferred to a vacuum flash jig at 120 mTorr for 30 s followed by 1.5 Torr for 15 s. This is then followed by an annealing step at 120 °C on a hotplate for 20 min. The sample is prepared in a glovebox.

**Steady-state PL and TRPL measurements of the triple cation perovskite sample.** The perovskite sample is sealed in a Linkam stage in a glovebox for PL and TRPL measurements. Before each measurement, the stage is purged with nitrogen. The steady-state PL measurement is performed by an IMA Photon etc. hyperspectral imaging system, and the excitation wavelength of the laser is 532 nm. The TRPL measurement is performed by an iHR 320 system (Horiba) with the excitation wavelength of 477 nm. Two neutral density filters are used to create low-level injection (incident laser energy $E_p$=22 nJ/cm$^2$) and high-level injection ($E_p$=1960 nJ/cm$^2$) conditions for the TRPL measurement. After steady-state PL and TRPL measurement, the sample is heated at 85 ºC for 36 hours. This purge-measurement-heating process is repeated 4 times. Throughout the whole process, the sample is placed inside the Linkham stage under the nitrogen environment.

**TRPL simulation.** TRPL is simulated by numerically solving a set of partial differential equations that describe carrier dynamics and recombination via finite-element methods (implemented by COMSOL v5.5). Here, a 1D model with 500 nm thickness is constructed for the perovskite.

Specifically, first, the perovskite is doped (p-type) with a dopant concentration of $N_d$ via complete ionization. The carrier density at equilibrium is then determined by the charge neutrality condition ($n_0$ for electron density and $p_0$ for hole density) and the Fermi level is determined self-consistently.

Second, excess carriers are generated by a laser pulse (wavelength $\lambda_{ex}$=477 nm, pulse width is 0.1 ns) with a density distribution along the 1D model described by the Beer-Lambert law

$$n_{ex} = p_{ex} = \frac{\alpha \lambda_{ex} E_p}{hc} e^{-\alpha x}, \tag{1}$$

where $h$ is the Planck's constant, $c$ is the speed of light, and $\alpha$=10$^5$ cm$^{-1}$ is the optical absorption coefficient of the perovskite at the wavelength of 477 nm. $E_p$ is the incident laser pulse energy that determines the injection level. In the following steps, $n = n_0 + n_{ex}$ is the total electron density and $p = p_0 + p_{ex}$ is the total hole density.

Third, the dynamics of all carriers is described by the drift-diffusion model

$$\frac{\partial n}{\partial t} = \nabla \cdot (\mu_n k_B T \nabla n + n \mu_n \nabla \phi) - R_r - R_A - R_{tn}, \tag{2}$$

$$\frac{\partial p}{\partial t} = \nabla \cdot (\mu_p k_B T \nabla p - p \mu_p \nabla \phi) - R_r - R_A - R_{tp}, \tag{3}$$

where $\mu_n$ and $\mu_p$ are the electron mobility and hole mobility, $k_B$ is the Boltzmann's constant, and $T$ is the temperature. $\phi$ is the electric potential inside the perovskite determined by the Poisson's equation

$$-\frac{\varepsilon}{q}\nabla^2\phi = p - n - N_d + N_{tc}, \tag{4}$$

where $\varepsilon$ is the permittivity of the perovskite, $q$ is the elementary charge, and $N_{tc}$ is the charged defect density described in Eq. (12). $R_r$ is the radiative recombination rate

$$R_r = B_r\left[np - N_cN_v\exp\left(-\frac{E_g}{k_BT}\right)\right], \tag{5}$$

where $B_r$ is the radiative recombination coefficient, $E_g$ is the bandgap, and $N_c$ and $N_v$ are the effective density of states for the conduction and valance band, respectively. $R_A$ is the Auger recombination rate

$$R_A = (C_{An}n + C_{Ap}p)\left[np - N_cN_v\exp\left(-\frac{E_g}{k_BT}\right)\right], \tag{6}$$

where $C_{An}$ and $C_{Ap}$ are the Auger recombination coefficients for electrons and holes, respectively. Here, we set $C_{An}=C_{Ap}=C_A$ for simplicity.

In this work, we describe the trap-assisted recombination by including four fundamental processes: (1) electrons in conduction band are captured by defects, (2) electrons in defect states are released to conduction band, (3) holes in valance band are captured by defects, and (4) holes in defect states are released to valance band. These four processes contribute to the trap-assisted recombination rates of electrons and holes and the dynamic occupation of defects[15, 30].

$R_{tn}$ and $R_{tp}$ in Eq. (2) and (3) are the trap-assisted recombination rates for electrons and holes

$$R_{tn} = N_t\sigma_nv_n[n - f_t(n + n_1)], \tag{7}$$

$$R_{tp} = N_t\sigma_pv_p[f_t(p + p_1) - p_1], \tag{8}$$

where $N_t$ is the defect density, $\sigma_n$ and $\sigma_p$ are the capture cross-section for electrons and holes, and $v_n$ and $v_p$ are the thermal velocity for electrons and holes. $f_t$ is the occupancy of the defects. $n_1$ and $p_1$ are defined as the thermal free electron and hole density when the Fermi level equals the defect energy level $E_t$,

$$n_1 = N_c \exp\left(\frac{E_t - E_c}{k_B T}\right), \tag{9}$$

$$p_1 = N_v \exp\left(\frac{E_v - E_t}{k_B T}\right), \tag{10}$$

where $E_c$ and $E_v$ are the conduction band edge and the valance band edge. Because electrons and holes are consistently captured and released by the defects, $R_{tn}$ and $R_{tp}$ must satisfy the dynamic equation of defect occupancy

$$N_t \frac{\partial f_t}{\partial t} = R_{tn} - R_{tp}. \tag{11}$$

The occupied defect then contributes to the charge in the perovskite

$$N_{tc} = -f_t N_t. \tag{12}$$

Since the perovskite sample is placed on the glass under the nitrogen environment in experiment, we consider the surface recombination is negligible in simulation and set zero flux at the boundaries.

Lastly, the simulation outputs the TRPL signal by integrating the radiative recombination over the entire material.

All parameters that are fixed in TRPL simulation is provided in Supplementary Table S1. Because the perovskite material in experiment is placed in nitrogen on a glass substrate, the front and back surface of the perovskite are considered to have very small surface recombination. In simulation, we use the Shockley-Read-Hall model and set a very small surface recombination velocity (1 cm/s) to define the boundary conditions. In addition, the electrical contacts at the boundaries are set as insulator with a voltage of 0 V.

The generic kinetic model used in Fig. 1, which is also known as the rate equation or the ABC model in literature, is simulated based on the following equation

$$\frac{dn}{dt} = -An - Bn^2 - Cn^3, \tag{13}$$

where A, B, and C are the fitting parameters which parameterize the trap-assisted recombination, radiative recombination, and Auger recombination, respectively.

**Correlation of $N_t$ and $\sigma$.** Eq. (7) and (8) can be re-written as

$$R_{tn} \approx aC_1(1 + C_2 e^{-\sigma C_4 t}), \tag{14}$$

$$R_{tp} \approx aC_1(1 - C_3 e^{-\sigma C_4 t}), \tag{15}$$

where $a=N_t\sigma$, $C_1 = \frac{vnp}{n+p}$, $C_2 = \frac{n_b-(n_b+p_b)f_b}{p}$, $C_3 = \frac{n_b-(n_b+p_b)f_b}{n}$, $C_4 = v(n+p)$. The subscript $b$ indicates the begging of TRPL and $f_b$ is the occupancy of the defects ($f_t$) at the begging of TRPL. In this derivation, we apply the approximation of $n \gg n_1$ and $p \gg p_1$ to Eq. (7) and (8) for simplicity of the discussion, which is valid when the dopant densityis not trivial. Releasing this approximation does not change the form of Eq. (14) and (15). It only modifies $C_1$, $C_2$, $C_3$, and $C_4$. By substituting Eq. (7) and (8) into Eq. (11) and integrating the equation, we have

$$f_t \approx \frac{n-[n_b-(n_b+p_b)f_b]e^{-\sigma C_4 t}}{n+p}. \tag{16}$$

Substituting this equation into Eq. (7) and (8) yields Eq. (14) and (15).

The second term in the bracket of Eq. (14) and (15) is a function of $\sigma$ only and is smaller than 1 when $C_2$ and $C_3$ are finite and $\sigma C_4 t$ is large. Eq. (14) and (15) then become proportional to $a$. When this condition is valid, the choice of $\sigma$ only marginally affect $R_{tn}$ and $R_{tp}$, and hence the parameter extraction, when $\sigma$ is greater than a certain value (in **Fig. 3**h, this value is around $10^{-14}$ cm$^2$).

Under the HLI condition, where the excess carrier density is large (and hence, $n+p$ is large), this condition is easier to satisfy. As the excess carrier density decreases under the LLI condition, the second term approaches 1, although it can still be small. This gives us a possibility to extract both $a$ and $\sigma$, where $a$ is easier to be extracted than $\sigma$, particularly under the LLI condition.

**Bayesian optimization.** The Bayesian optimization algorithm starts with 4 TRPL simulations in which the parameters of interest take randomly selected values from the defined ranges. We then calculate the difference between the simulated TRPL and the target TRPL using a cost function. Based on these evaluations, the algorithm starts to fit the Gaussian Process Model and update it with the simulation result using the Bayesian methods (Fig. 1a). In the fitting of the Gaussian Process Model, a covariance kernel function called ARD (automatic relevance determination) Matern 5/2 kernel is used. Based on the updated Gaussian Process Model, we use the acquisition function of expected improvement to find the next best parametric values possible for TRPL simulation. The new data produced by the simulation is fed back to the algorithm to update the Gaussian Process Model. This process is an iterative loop. The algorithm continues for 600 iterations/tests and outputs all tested parametric values and cost values in a table. This table is re-arranged by small cost value. Then we create a sub-table which contains the tests with cost value smaller than $10^{-3}$+the smallest cost. If the number of eligible tests in the sub-table is greater than 10, we take the average of the tested parametric values as the extracted parameters and we calculate the standard deviation of these values as the uncertainty. If the number of eligible tests is smaller than 10, the tested parametric values with the smallest cost value are considered as the extracted parameters, and the standard deviation of the top 10 tests gives the uncertainty. This algorithm is implemented by MATLAB.

We design 8 cost functions as follows.

$$f_0 = \frac{1}{y_{end}}\sqrt{\sum_i (h_i - y_i)^2}, \tag{17}$$

$$f_1 = \log\left[\frac{1}{y_{end}}\sqrt{\sum_i (h_i - y_i)^2}\right], \tag{18}$$

$$f_2 = \frac{1}{\log y_{end}}\sqrt{\sum_i (\log h_i - \log y_i)^2}, \tag{19}$$

$$f_3 = \log\left[\frac{1}{\log y_{end}}\sqrt{\sum_i (\log h_i - \log y_i)^2}\right], \tag{20}$$

$$f_4 = \sqrt{\sum_i \left(\frac{h_i - y_i}{y_i}\right)^2}, \tag{21}$$

$$f_5 = \log\left[\sqrt{\sum_i \left(\frac{h_i - y_i}{y_i}\right)^2}\right], \qquad (22)$$

$$f_6 = \sqrt{\sum_i (h_i - y_i)^2}, \qquad (23)$$

$$f_7 = \log\left[\sqrt{\sum_i (h_i - y_i)^2}\right], \qquad (24)$$

where $h_i$ and $y_i$ are the simulated and experimental/target TRPL at the time point $i$. We use $f_7$ for HLI only fitting and $f_5$ when not specified. The normalization operation in $f_5$ (and $f_4$) averages the weight of fitting at all time points in a TRPL curve. In contrast, when using $f_7$, the high PL intensity at the beginning of TRPL intrinsically leads to a better fitting at the beginning than at the end. This makes $f_7$ particularly suitable for HLI only fitting because the radiative and Auger recombination show large effect at the beginning of TRPL. Note that, although $f_0$ (and $f_1$) give the same result as $f_6$ (and $f_7$) during LLI only or HLI only fitting, they differ in the fitting to both LLI and HLI. Here we omit the square root of $N$, which is present in the expression of RMSE, in all cost functions because it scales the cost values for all iterations/fits, including the good and bad ones, by the same amount and does not affect the parameter extraction.

The logarithm-type cost function is designed for two reasons. Firstly, it increases the probability for BO to find the target (global optimum) from a group of values that can produce similar results (local optima). This is because the logarithm operation unevenly scales the good and bad fits, i.e., the cost value is significantly lower for the good fit than the bad ones. Specifically, BO assesses the probability using an acquisition function where a large function value indicates a higher probability. Here we use the acquisition function of expected improvement, i.e., the expected value of max[0, $f_{best}$–g($x$)] where $f_{best}$ is the best/minimum cost observed so far and g($x$) is the Gaussian process model of the cost[31]. The logarithm operation on the RMSE gives a theoretically negative infinite cost, and hence a very large value of $f_{best}$–g($x$), when $x$ equals the global optimum, whereas the local optima give finite and often small values of $f_{best}$–g($x$). A second reason to use a logarithm-type cost function is to re-scale the cost of different parameters to the same order of magnitude. This can help BO to avoid focusing only on optimising the parameters that have high costs during multi-parameter extraction.

Here we compare the performance of BO for 5 different kernel functions used in a single-parameter extraction of $N_t$. ARD Matern 5/2 and ARD Matern 3/2 outperform ARD exponential, ARD Squared exponential, and ARD rational quadratic (Supplementary Fig. S11), which is consistent with literature[32]. Note that, when TRPL experiment data is very noisy, the cost values could increase, particularly for a good fit, i.e., a good fit may have a cost value that is close to bad fits. This means noisy experimental data can induce difficulty to find the best-fit parameters. In Gaussian process, this could mean adding additional noise to the covariance function for the prior Gaussian distribution. In this work, the covariance function already includes the contribution from the kernel function and a Gaussian noise with its own variance. This variance and the kernel parameters are found during fitting. This could help to reduce the difficulty when the noise in data is Gaussian. For more complicated noise, this will be an interesting topic for future comprehensive studies.

**Data availability**

The program for parameter extraction can be found at GitHub[33], with a manual at Zenodo[34]. Data for this paper are available at Zenodo[35].

**Author contributions**

H.Z. conceived the idea, performed the machine learning study, and prepared the manuscript. V.A. and G.T. synthesized the perovskite samples. H.Z., A.M., A.D.B., and Z.L. performed the TRPL and PL measurements. K.C. and T.W. supervised the project. All authors contributed to discussion and manuscript revision.

**Conflict of Interests**

There are no conflicts of interest to declare.

**Acknowledgements**

The work is supported by the Australian Centre for Advanced Photovoltaics (ACAP) and received funding from the Australian Renewable Energy Agency (ARENA). H.Z. acknowledges the support of the ACAP Fellowship. H.Z. thanks Pawsey for providing the Nimbus Research Cloud Service.